\documentclass[runningheads]{llncs}
\titlerunning{UMCL}
\usepackage[T1]{fontenc}
\usepackage{graphicx}
\usepackage{booktabs}
\usepackage{multicol}
\usepackage{multirow}
\usepackage{cite}
\begin{document}
\title{Unified Medical Image-Text-Label Contrastive Learning With Continuous Prompt}
\author{Yuhao Wang\inst{1}}

\authorrunning{anonymous et al.}
%
\institute{Beijing University of Posts and Telecommunications\\
\email{wangyuhao@bupt.edu.cn}
}

\maketitle              

\begin{abstract}

Contrastive language-image Pre-training (CLIP) \cite{CLIP} can leverage large datasets of unlabeled Image-Text pairs, which have demonstrated impressive performance in various downstream tasks. Given that annotating medical data is time-consuming and laborious, Image-Text Pre-training has promising applications in exploiting large-scale medical image and radiology report datasets. However, medical Image-Text Pre-training faces several challenges, as follows: (1) Due to privacy concerns, the amount of available medical data is relatively small compared to natural data, leading to weaker generalization ability of the model. (2) Medical images are highly similar with only fine-grained differences in subtleties, resulting in a large number of false-negative sample pairs in comparison learning. (3) The hand-crafted Prompt usually differs from the natural medical image report, Subtle changes in wording can lead to significant differences in performance. In this paper, we propose a unified Image-Text-Label contrastive learning framework based on continuous prompts, with three main contributions. First, We unified the data of images, text, and labels, which greatly expanded the training data that the model could utilize. Second, we address the issue of data diversity and the impact of hand-crafted prompts on model performance by introducing continuous implicit prompts. Lastly, we propose a Image-Text-Label contrastive Training to mitigate the problem of too many false-negative samples. We demonstrate through sufficient experiments that the Unified Medical Contrastive Learning (UMCL) framework exhibits excellent performance on several downstream tasks.


\keywords{Multi-Modal Pre-training \and Medical Vision-and-Language \and Continuous Prompt \and Contrastive Learning} 
\end{abstract}
\section{Introduction}

Visual-Language models\cite{Align,2022UnifiedCL,COOP,OscarOA,1VinVLRV} aim to learn generic visual representations through Image-Text pairs. With the development of multimodal healthcare AI \cite{MultimodalbiomedicalAI}, increasing amounts of multimodal healthcare data are becoming available, including images, text, and  other multimodal data. These massive amounts of unlabeled data provide a good foundation for the application of self-supervised learning methods in healthcare. One such method is Contrastive Language-Image Pre-training (CLIP) \cite{CLIP}, which can be pre-trained using a large number of unlabeled Image-Text pairs. CLIP has demonstrated impressive performance in a variety of downstream tasks, e.g. zero-shot classification, cross-modal retrieval.

In the medical field, where data annotation is often time-consuming and laborious\cite{Expert}, the superior performance of Image-Text Pre-training in zero-shot classification can enable the diagnosis of rare diseases with zero or few shots. However, existing Image-Text Pre-training has several issues in the medical field:
1. Medical image reports contain rich patient information, which may result in patient privacy leakage. The insufficient existing medical image-report text pairs cannot provide the same powerful generalization ability to Image-Text Pre-training models as natural images.
2. The hand-crafted Prompt usually differs from the natural medical image report, Subtle changes in wording can lead to significant differences in performance
3. Medical images are usually highly similar, differing only in subtleties, and there are often inextricable links between unpaired images and reports, which leads to a large number of false-negative samples in the traditional Pre-training paradigm.

For all the reasons above, only a few studies have been conducted on Vision-language pretrain in the medical field\cite{ShihChengHuang2021GLoRIAAM,MedCLIP,Convirt,HongYuZhou2021GeneralizedRR,ZhihongChen2022MultiModalMA}.
\cite{HongYuZhou2021GeneralizedRR} adopted cross-supervised paradigm combining contrast learning and image captioning is used to facilitate the adoption of generic representation capabilities for image encoders.  \cite{ZhihongChen2022MultiModalMA}  borrows the idea of MAE to perform graphical Pre-training by cross-modal mask reconstruction, so that the trained model can support MedQA. However, all these Image-language pretrain cannot support zero-sample inference as flexibly as the CLIP Pretraining paradigm. Most similar to CLIP Pretrain paradigm, ConviRT \cite{Convirt} uses Pre-training by bi-directional comparison between image text targets using paired text data to compute different InfoNCE \cite{InfoNCE} losses separately, while GLoRIA \cite{ShihChengHuang2021GLoRIAAM} proposes an attention-based framework by comparing image sub-regions and words in paired reports to learn global and local representations.

To address existing issues, we propose a unified Image-Text-Label Pre-training framework based on continuous prompts.
Firstly, by constructing a prompt, we map the Image-Label dataset and the Image-Text dataset into the same implicit space. This greatly expands the number of datasets that can be used for Pre-training, enabling multiple sources and types of datasets to be trained in a unified framework. By using different types of supervised signals in the model, it has both the excellent generalization of the Image-Text Pre-training model and the characterization ability of robust transfer learning.
Secondly, Image-Text-Label contrastive Training is proposed to solve the problem of false negative samples in the contrast learning process.
Finally, we introduce continuous prompts, which effectively avoid the performance difference brought by hand-crafted prompts and bridge the difference between hand-crafted prompts and natural medical image reports. This improves the ability of zero-shot inference.
Extensive experiments show that our unified Image-Text-Label Pre-training framework can effectively exploit both types of data and achieve excellent performance in multiple downstream tasks, such as zero-shot inference, transfer learning, and cross-modal retrieval.
\section{Method}

Our proposed UMCL model, as shown in Fig. 1, utilizes continuous prompts to construct corresponding prompts on Image-Label data. This allows disease label to be mapped to a unified feature space while pretraining. To address the issue of false negatives in the Image-Text Pre-training model, our proposed Image-Text-Label contrastive Training can effectively leverage the specific representational information in the Image-Label data, as well as the generic representational information in the Image-Text pair. This enables the learned model to achieve superior performance with zero-shot classification and transfer learning

\begin{figure}[t]
    \centering
    \includegraphics[width = 1\textwidth]{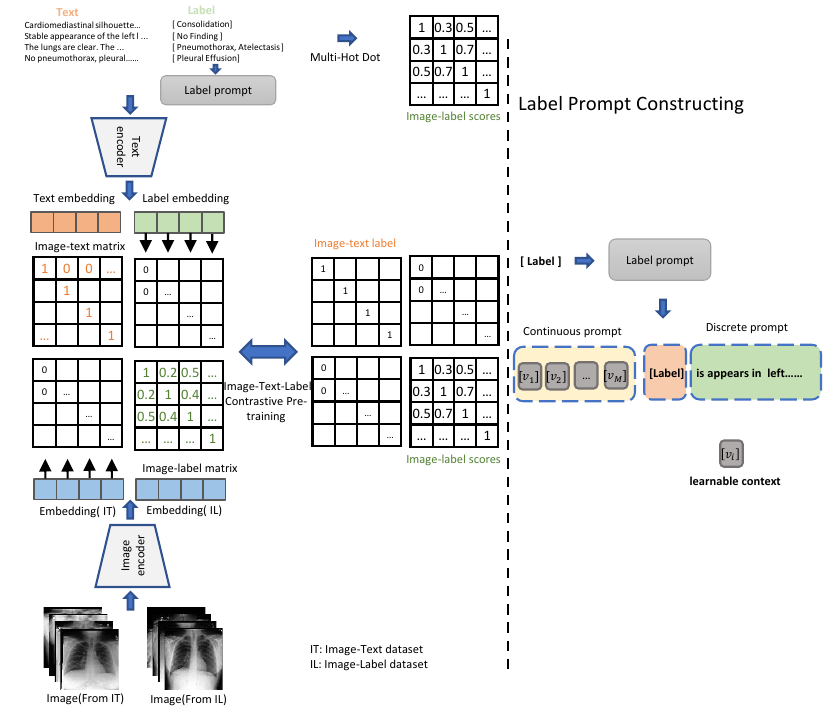}
    \caption{Overview of UMCL framework. The left part is the UMCL framework and the right part is the continuous prompt constructing process of Image-Label dataset.}
    \label{fig:2}
\end{figure}

\subsection{Vision and Text Encoder}
For the image encoder, we utilized the current state-of-the-art image encoder architecture SwinTransformer\cite{ZeLiu2021SwinTH}, to encode the image as a visual vector with a fixed dimension. Regarding the text encoder, we adopted the previously established approach of using BioClinicalBERT\cite{bioclinicalbert} and encoded the corresponding text as a text vector with the same dimension as the visual vector.

\subsection{Unified Image-Text-lable pretrain with continuous prompt}
Inspired by \cite{JieLiu2023CLIPDrivenUM},to address the issue of insufficient data amount for the Image-Text Pre-training model, we incorporated the common Image-Label dataset in the model training. We unified the Image-Text and Image-Label datasets by constructing a prompt that maps Label into the contrast learning paradigm. However, we focused on the differences between the natural radiology report and prompts constructed from the label for the two types of sentences, originating from different data types, differentiating them at the input level. Specifically, for the Image-Label data, we first followed the relevant disease description template proposed in GLoria \cite{ShihChengHuang2021GLoRIAAM} to create a textual description prompt (discrete prompt) for the disease label. This template was reviewed by several radiologists and contains a reasonable and rich description of the disease, lesion condition, severity, and location, which can reduce the difference between image labels and natural medical image reports to some extent. 

Subsequently, inspired by \cite{P-Tuning,COOP} we proposed a continuous image labeling-based medical imaging system and characterized the Image-Label dataset using a continuous prompt approach. Continuous prompt is a set of learnable vectors inserted for the generated text description cues that, unlike hand-crafted Prompt, do not have to be bound to a deliberate vocabulary but use an invisible unified vector. 

During model training, continuous prompts can be trained end-to-end, where the parameters can be updated during the Pre-training process. Our framework unifies Image-Text-Label pretraining, and continuous prompt can effectively capture the commonalities and differences between the labels and the original medical image reports, bridging the differences in the Image-Label dataset in the downstream task of zero-shot classification for the Image-Text pretraining model. This formula can be expressed as follows:

\begin{equation}
\centering
    T=[\mathrm{V}]_1[\mathrm{~V}]_2 \ldots[\mathrm{V}]_M[\mathrm{CLASS}][\mathrm{F}_{ds}(\mathrm{CLASS})]
\end{equation}
Where$[\mathrm{V}]_m(m \in\{1, \ldots, M\}$ is a vector with the same dimension as word embeddings, and M is a hyperparameter specifying the number of context tokens. $\mathrm{F}_{ds}$ denotes the process of discrete prompt construct, which is to search several common hand-crafted sentences by class names. For the Image-Text data, we adopted the same preprocess pipeline as CLIP.

\subsection{Image-Text-Label contrastive Training}

Traditional Image-Text Pre-training models typically encounter the problem of excessive false negatives. This problem arises because image reports from different medical images are unlikely to be completely different. As a result, the effectiveness of the original CLIP model, applied directly to medical Image-Text report pairs datasets, is limited, as the model often forces the separation of otherwise semantically consistent image reports. Our Image-Text-Label Pre-training framework effectively introduces the disease labels of images, allowing the disease labels of images to effectively modeling the consistency of images and corresponding prompts at the semantic consistency level. During the model training process, our successive prompts can be automatically optimized so that the final image labels constitute a prompt highly close to the natural medical image reports. 

For the Image-Label dataset, the similar score of unparired images and prompts constructed from disease label is  calculated by:
\begin{equation}
y_{i,j}=\frac{\mathbf{l}_{i}^{\top} \cdot \mathbf{l}_{j}}{\left\|\mathbf{l}_{i}\right\| \cdot\left\|\mathbf{l}_{j}\right\|}
\end{equation}
where $y_{i,j}$ denotes the similar score of $i$ images and $j$ prompts constructed from disease label, and l denotes the multi-hot vectors generated by disease labels, e.g. ['consolidation','lung Opacity'] is denoted as $[0,1,.....1,0]$.

The predicted similar scoreis also obtained by L2 normalize:

\begin{equation}
    s_{i,j}=\mathbf{v}_{i}*\mathbf{t}_{j}^{T}
\end{equation}
where ${v}_{i}$ and ${t}_{j}$ represent the image embedding and text embedding.

\begin{equation}
s
_{i j}=\frac{s i j}{\sqrt{\sum_{i=1}^N \sum_{j=1}^N s_{i j}^2}}
\end{equation}

s indicates the medical semantic similarity. We adopt the L2 to nomralize the predicted similar score, and 
Image-Text-Label contrastive Training can be formulated as followings:
\begin{equation}
\mathcal{L}=-\frac{1}{N_{b a t c h}} \sum_{i=1}^{N_{b a t c h}} \sum_{j=1}^{N_{b a t c h}} y_{i j} \log {s}_{i j}
\end{equation}

For the image-text dataset, Due to the lack of semantic similarity index between different samples, we only assign the similarity labels of pair samples as 1, while the similarity labels of un-paired samples is 0.

More specifically, we outlaw the softmax layer in Cross Entropy loss and utilize L2 normalization, which avoids the original contrast learning loss from forcibly maximizing the distance between un-paired samples when faced with image-text datasets lacking semantic similarity labels during model training. For the Image-Text data, the loss can optimize only the Image-Text sample pairs in pair data and ignore other datasets without similar soc because there is no structured disease label. For the Image-Label, the loss function can effectively constraint the similarity between the image and the corresponding prompt by the disease label. In this way, the loss function can effectively unify different Pre-training data sources in the Image-Text-Label dataset, allowing the model to be equiped with a more discriminative feature representation in downstream transfer learning tasks and to learn generic features.

\section{Experiment}
\subsection{Datasets and Implementation details}

\subsubsection{Pre-train datasets}
\ 
\newline
 \indent \textbf{CheXpert}\cite{CheXpertAL}:CheXpert is a publicly available Pre-training dataset developed by the Stanford University School of Medicine. This dataset contains a vast collection of 224,316 chest X-rays obtained from 65,240 patients. It includes labeled information from free-text reports describing the presence of 14 common chest diseases and findings in the images, such as pneumonia, nodules, cardiopulmonary enlargement, and more.

\textbf{MIMIC-CXR}\cite{AMIMIC}: MIMIC-CXR (Medical Information Mart for Intensive Care - Chest X-ray) is another Pre-training dataset containing nearly  227,835 chest X-ray images, each of which corresponds to a text report from various US medical institutions. 

\subsubsection{Evaluation Datasets}
\ 
\newline
 \indent \textbf{CheXpert-5x200}: we fellow the the config of GloRIA\cite{ShihChengHuang2021GLoRIAAM}, sample a multi-class classification dataset from the testing split.The class of heXpert-5x200 include Atelectasis, Cardiomegaly, Edema, Pleural, Effsion, and each class contain 200 positive samples.

\textbf{MIMIC-5x200}: For evaluation, we also sample a MIMIC-5x200 dataset for the same five tasks above.

\textbf{COVID-19}\cite{COVID-19}:This dataset was collected from several hospitals in QATAR, containing approximately equal numbers of COVID and non-COVID tags cases. We sample a balanced subset for COVID-19 and Non-COVD19 1:1 for 2000 samples.

\textbf{RSNA pneumonia} \cite{RSNA-pneumonia}: The RSNA pneumonia dataset, published by the Radiological Society of North America (RSNA) in 2018, containing 26,684 chest X-ray images of patients with pneumonia and normal people of different ages, genders, and races. We fellwing the configs in \cite{MedCLIP},and sample a balanced subset for pneumonia and non-pneunomia and use it for evaluation.

\subsubsection{Implementation details}

We utilize a multimodal Pre-training model that combines ViT architecture\cite{ZeLiu2021SwinTH} as the image encoder and bioclinical bert\cite{bioclinicalbert} architecture as the text encoder. Our model involves merging the initial hand-crafted prompt with a continuous prompt in the embedding layer of the text encoder. The continuous prompt has a length of 32, and we set the specific dimension to 512 learnable vectors. Additionally, we set the implicit dimension of the encoder to 768. For all Pre-training experiments, we train the model for 100,000 steps using the Adam optimizer with a learning rate of 1e-5. We also set the pre-warming ratio to 10 and use a linear learning rate scheduler after the pre-warming. The pre-processing steps involve scaling images to 256 × 256 and uniformly processing text to a length of 77 tokens. We complete the training on an NVIDIA A40 GPU, which offers fast and efficient processing capabilities.

\subsection{Results and Discussions}
\subsubsection{zero-shot classification}
Following previous work, we we evaluate zero-shot image classification on four datasets, CheXpert-5x200, MIMIC5x200, COVID and RSNA to assess the generalizability of our model. We represent the image labels by constructing a continuous prompts , and comparing the similarity of image embedding and text embedding . The results are presented in Table \ref{tab:1}. We adopted the ACC(accuracy) as the evaluation metric. The results show that our method exhibits excellent performance on the zero-shot task for multiple datasets. Of interest is that, unlike other models, the insertion of our continuous prompt largely enhances the advantages of prompt ensemble, which indicates that the contextual vectors learned during training can well capture the differences and commonalities between hand-craft prompts and natural medical image reports, thus unifying the three feature spaces of Image-Text-Label.
In particular, we achieve excellent performance on both COVID and RSNA datasets, and our model never sees the COVID and RSNA datasets during training, while the ACC of zero-shot on the two datasets reaches 0.748 and 0.764, respectively, which demonstrates the strong generalization ability of our model

\begin{table}[]
\vspace{-1.0em}
    \centering
\setlength{\tabcolsep}{3mm}{} 
\begin{tabular}{c|cccc}
   \toprule
  ACC & CheXpert-5x200 & MIMIC-5x200 & COVID  & RSNA\\
   \midrule
    CLIP& 0.201 & 0.191 &  0.506 &  0.498 \\
   CLIP-ENS& 0.203 & 0.225 &  0.509 &  0.505 \\
   \hline

   ConVIRT& 0.418 & 0.401 &  0.518 &  0.473 \\
   ConVIRT-ENS& 0.422 & 0.401 &  0.664 &  0.464 \\
   \hline

   GLoRIA & 0.432 & 0.330 &  0.709 &  0.580 \\
   GLoRIA-ENS& 0.421 & 0.338 &  0.570 &  0.475 \\
   \hline

   UMCL &  0.527 & 0.497 &  0.707 &  0.670 \\
   UMCL-ENS&\textbf{0.541}  & \textbf{0.505} &  \textbf{0.748} &  \textbf{0.764}\\

   \bottomrule
\end{tabular}

    \caption{Results of Zero-shot classification on several  CheXpert-5x200, MIMIC5x200, COVID and RSNA, suffix ENS represents Prompt-Ensemble.}
    \label{tab:1}
\vspace{-4.0em}
\end{table}

\subsubsection{Classification with finetune}
We validated the strong transfer learning ability of our model by fine-tuning it on same datasets as above. The results are shown in Table \ref{tab:2}. In comparison to other models, ours demonstrated better performance after fine-tuning, suggesting that our model effectively learns more discriminative representations during the training process. Meanwhile, compared with Table 1 \ref{tab:1}, it can be found that the performance difference between finetung and zero-shot is lesser margin. This demonstrates that our model has strong general and specific representations at the same time, and indicates the promising application of our Image-Text-Label Pre-training model in scenarios with limited labeled data. 

\begin{table}[]

\setlength{\tabcolsep}{3mm}{} 
    \centering
    \begin{tabular}{c|cccc}
    \toprule
        ACC & CheXpert-5x200 & MIMIC-5x200 & COVID & RSNA\\
        \hline
        ImageNet & 0.32& 0.283 & 0.602 & 0.756\\
        CLIP & 0.302  & 0.278&0.586 &0.730\\
        ConVIRT& 0.477&0.404 & 0.698 &0.784\\
        GLoRIA& \textbf{0.537}
        &0.359 &0.762 & 0.783\\
        \hline
        UMCL& 0.532 &\textbf{0.545}& \textbf{0.773}&\textbf{0.803}\\
    \bottomrule
    \end{tabular}
    \caption{The ACC of finetue UMCL on  CheXpert-5x200, MIMIC5x200, COVID and RSNA}
    \label{tab:2}
\vspace{-2.0em}
\end{table}

\subsubsection{Cross-modal Retrieval}
As with previous work, we evaluated our proposed model using the CheXpert-5x200 dataset. Specifically, we computed several Image-Text pairs with the largest similarity scores by encoding the images and corresponding sentence embeddings. We then evaluated our model's accuracy using Precision@K to separately compute the accuracy in top K retrieval reports/sentences. The specific results are presented in Table \ref{tab:3}, which demonstrate that our Image-Text-Label Pre-training model achieved optimal results on the metrics of multiple sampled samples. 

\begin{table}[]
\vspace{-2.0em}
\setlength{\tabcolsep}{3mm}{} 
    \centering
    \begin{tabular}{c|cccc}
    \toprule
        Model & P@1 &P@2 & P@5 &P@10\\
        \hline
        CLIP & 0.211  &0.205 &0.217 & 0.195\\
        ConVIRT&0.205 & 0.217 & 0.209& 0.210\\
        GLoRIA& 0.447  &0.462& 0.464&0.457\\
        \hline
        UMCL& \textbf{0.467} &\textbf{ 0.472} &\textbf{0.480}&\textbf{0.467}\\
    \bottomrule
    \end{tabular}
    \caption{Cross modal Retrieval Results, we use P@1, P@2, P@5, P@10 to evaluate the cross-modal retrieval performance}
    \label{tab:3}
\vspace{-4.0em}
\end{table}

\subsubsection{Ablation study}
To investigate the impact of different length of the continuous prompt and the Image-Label dataset to our approach, we conduct a rigorous ablation study on Chexpert-5x200. when the length of continuous prompt is set to 16 or 64, the UMCL has a noticeable performance degradation. This demonstrate that the continuous prompt have ontly enhance efficiency, while the zero-shot classfication highly depends on the loading of discrete prompt. We then remove the continuous prompt and Image-Label data, the results show that the introduction of Image-Label dataset have a significant improvement on the model performance and the continous prompt bridge the gap between the Text and the Label. In summary, our proposed components improve on all metrics, which demonstrate the prominent effectiveness of our approach.

\begin{table}[]%
\setlength{\tabcolsep}{3mm}{} 
    \centering
    \begin{tabular}{c|c}
    \toprule
        setting& CheXpert-5x200 \\
        \hline
        context-16 & 0.532\\
        context-32 &\textbf{0.541}   \\
        context-64& 0.521 \\
        \hline
        UMCL &\textbf{0.541} \\
        w/o context& 0.497 \\
        w/o label-data & 0.201\\
    \bottomrule
    \end{tabular}
    \caption{Ablation study}
    \label{tab:2}
\end{table}

\subsubsection{T-SNE Visulization}
Finally, we compared the t-SNE results of our Image-Text-Label Pre-training model and the CLIP model. As shown in Figure \ref{fig:2}, the image encoder of the CLIP model was found to be less effective in distinguishing different chest films, while our model was able to effectively distinguish between multiple diseases. It is worth noting that our model achieved a significant clustering effect for different diseases.

The results suggest that our model outperforms the image encoder of the CLIP model in this specific task. The ability to differentiate between various diseases is essential in medical imaging, and our model's success in achieving this could be of great value in clinical settings.

\begin{figure}[t]
    \centering
    \includegraphics[width = 1\textwidth]{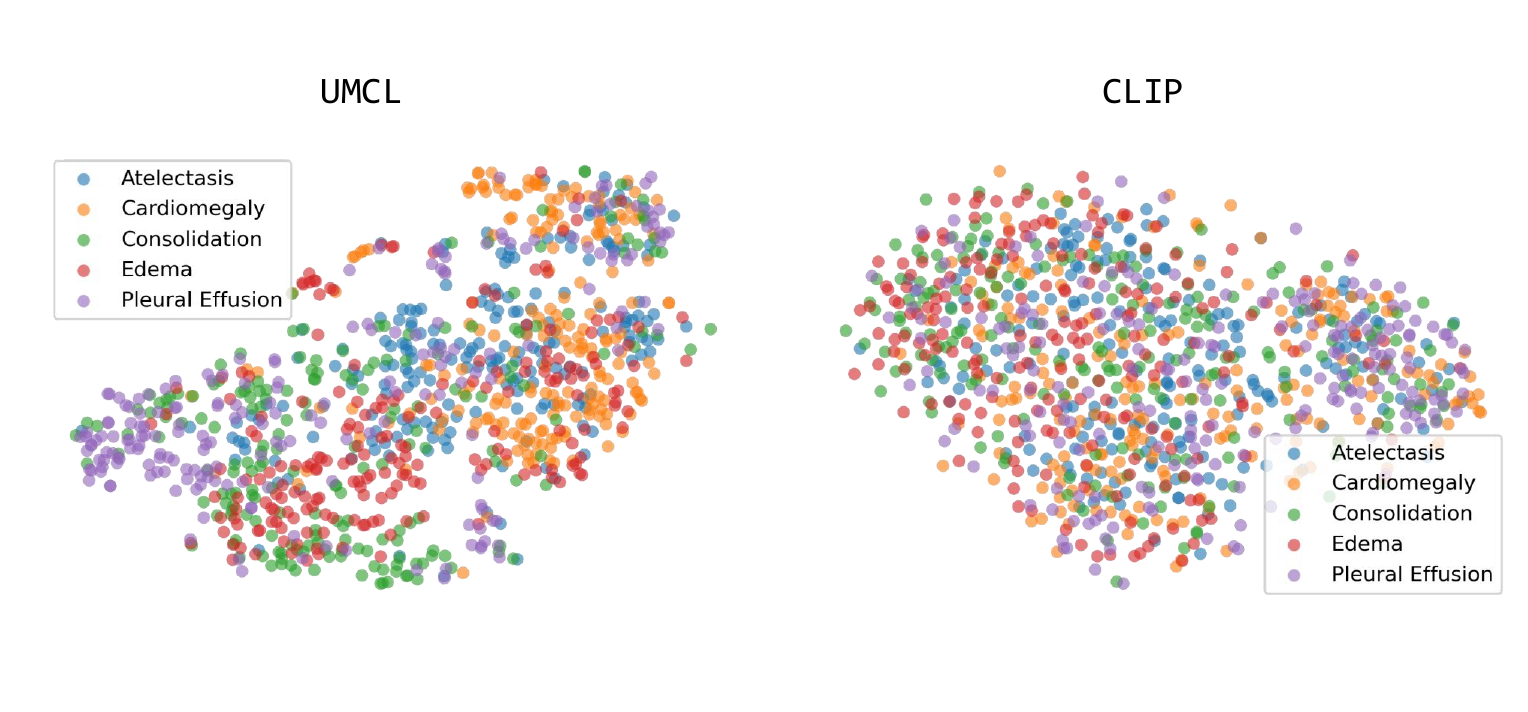}
    \caption{The Image embedding of UMCL and CLIP on CheXpert-5x200, Dimension reduced by t-SNE.}
    \label{fig:2}
\end{figure}

\section{Conclusion}
In this paper, we introduced a new Pre-training paradigm, which unifies the Image-Text-Label modality using continuous prompts and a Image-Text-Label contrastive Training function. Our proposed model effectively integrates three different data types into the same feature space, thus addressing several issues related to Image-Text-text Pre-training models, such as the lack of sufficient data, the strong dependence on hand-crafted prompts, and the high false-negative rate.

Our comprehensive experimental evaluation demonstrates the effectiveness of our approach in various tasks, including zero-shot classification, cross-modal retrieval, and fine-tuned classification, across multiple datasets. Our results indicate that our method outperforms existing state-of-the-art Pre-training models, demonstrating the potential of our approach in various applications. 

\subsection*{Acknowledgements}
%
%
%
%

\bibliographystyle{splncs04}
\bibliography{mybibliography}

\end{document}